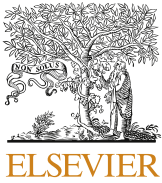
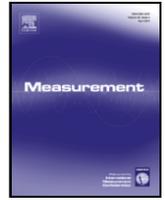

# Temperature compensation in high accuracy accelerometers using multi-sensor and machine learning methods

Lorenzo Iafolla [a],[1], Francesco Santoli [b], Roberto Carluccio [a], Stefano Chiappini [a], Emiliano Fiorenza [b], Carlo Lefevre [b], Pasqualino Loffredo [b], Marco Lucente [b], Alfredo Morbidini [b], Alessandro Pignatelli [a], Massimo Chiappini [a]

[a] *Istituto Nazionale di Geofisica e Vulcanologia, Via di Vigna Murata 605, 00143 Rome, Italy*
[b] *Istituto di Astrofisica e Planetologia Spaziali (IAPS), Istituto Nazionale di Astrofisica (INAF), Via del Fosso del Cavaliere 100, 00133 Roma, Italy*

ABSTRACT

Temperature is a major source of inaccuracy in high-sensitivity accelerometers and gravimeters. Active thermal control systems require power and may not be ideal in some contexts such as airborne or spaceborne applications.

We propose a solution that relies on multiple thermometers placed within the accelerometer to measure temperature and thermal gradient variations. Machine Learning algorithms are used to relate the temperatures to their effect on the accelerometer readings. However, obtaining labeled data for training these algorithms can be difficult. Therefore, we also developed a training platform capable of replicating temperature variations in a laboratory setting.

Our experiments revealed that thermal gradients had a significant effect on accelerometer readings, emphasizing the importance of multiple thermometers.

The proposed method was experimentally tested and revealed a great potential to be extended to other sources of inaccuracy as well as to other types of measuring systems, such as magnetometers or gyroscopes.

## 1. Introduction

This work was done within the framework of GAIN (*Gravimetro Aereo INtelligente*, Airborne Intelligent Gravimeter) project, which aimed to create a new strapdown gravimeter for airborne gravity surveys. The scope of gravity measurement is broad and generally related to understanding the structure of the Earth [21]. For example, gravimetry is one of the oldest methods for geophysical exploration [22] and for measuring changes in Earth structure over time [28].

Airborne gravimetry is a demanding method for geophysical surveying, as high-accuracy measurements are required in a very noisy environment [26]. State-of-the-art airborne gravimeters can reach down to 1mgal ($10^{-5}$m/s²) accuracy [1,9,13], and further improvements are needed to detect small sources of gravity anomalies. High sensitivity accelerometer and gravimeters can easily be more accurate than 1mgal, but within moving platforms, such as an aircraft, disturbances pose the major limitation. Examples of disturbances are temperature variations, unaccounted rotations of the reference frame, vibrations, etc. The conventional solution is using stabilizing platforms to actively (e.g., by means of motors and heaters) reduce them physically. However, space and power are scarce resources on aircrafts and on other moving platforms (e.g., unmanned aerial vehicles, autonomous underwater vehicles, space satellites). This represents a big limitation to using active stabilizing platforms.

In the GAIN project we avoided stabilizing platform in favor of a novel paradigm consisting of three pillars: multi-sensor system, Machine Learning (ML), and training platform. The multi-sensor system resembles a strapdown gravimeter, typically equipped with a tri-axial accelerometer, a tri-axial gyroscope, and a GNSS positioning system [17, 18]. The idea behind is that the multi-sensor system measures both the gravity and the disturbances. However, employing a multi-sensor system presents challenges, including the need to model it. Specifically, we need to retrieve the gravity from all the measured signals. Moving to the second pillar of our paradigm, the former challenges are addressed with ML algorithms that can learn from data (*training data*) how to model the system effectively. Still, a persistent challenge in ML is the requirement for labeled data (data with known desired output) for training. One potential solution involves obtaining them from a known environment, such as an area where the gravity field has been previously measured using alternative techniques. However, implementing this so-lution may be costly, time-consuming, or impractical, especially in spaceborne applications. Moreover, a complexity arises from the need for training data to cover various scenarios, such as different temperature ranges encountered in operational contexts. The third pillar tackles these challenges by utilizing a hardware platform, called training platform, to simulate the operational environment within a laboratory setting. This platform facilitates the acquisition of labeled data, effectively overcoming the aforementioned issues. The paradigm is implemented in four sequential steps. Initially, the multi-sensor system is deployed on the training platform to gather training data. Subsequently, the ML algorithm undergoes training and testing. In the third step, the multi-sensor system is installed onboard the moving platform for utilization in the operational context. Finally, the gravity measures are extracted from the multi-sensor output employing the previously trained ML algorithm.

Since measuring gravity is essentially equivalent to measuring acceleration (according to Einstein's equivalence principle), the concepts discussed in this paper concerning gravimeters are also applicable to accelerometers, and vice versa. Consequently, we extended the scope of our study from the narrower challenge of compensating for disturbances in gravimeters to the broader one related to accelerometers. Specifically, we addressed the challenge of compensating for temperature, which is a major source of inaccuracy in most applications. Temperature variations can arise from different sources, for instance in airborne gravimetry from the changing exposure to sunlight during the survey. These variations can affect the physical properties of the accelerometer by changing its sensitivity (i.e., the calibration factor) and thus modulating its output [3]. For example, if an accelerometer measures the gravity in a stationary environment, its output will reflect the temperature variations instead of being constant. Other effects of temperature variations are known, such as the differential thermal dilatation of the feet of a gravimeter, which changes its alignment to the vertical axis and, thus, its reading [7]. In other studies, thermal deformation of the package was identified and investigated as a major source of error [5].

Many strategies have been used for temperature compensation. For example, most of gravimeters and spaceborne accelerometers are thermally insulated and feature active thermal control systems [6,15,25,27]. Similarly, high accuracy MEMS accelerometers may also feature an active thermal control system [31]. In [19], a temperature gradient was induced to reduce the warpage of the bonded slices of a MEMS device. Other accelerometers were designed to be insensitive to temperature [12,20,30].

In GAIN, active thermal control systems were avoided in favor of temperature rejection in data post processing. This approach has been widely used and investigated. For example, spaceborne accelerometers typically feature a thermometer in addition to the active thermal control. Using a linear relationship, the residual effects of the temperature could be removed. However, determining an accurate mathematical relationship is required for achieving more accurate corrections. For this reason, several works have investigated methods to determine it. For example, in [29,32] the authors focused on developing fast and efficient methods for calibration while in [10,11] they have analyzed machine learning based methods. However, the former works did not explore thoroughly the impact of thermal gradients over the accelerometer performance. This aspect was discussed in [23], where a MEMS based accelerometer was exposed to thermal gradients. In the latter work, the authors concluded that thermal gradients are usually not relevant in most applications, as they are typically small in miniature MEMS devices. Our findings demonstrate that this conclusion does not apply in general, for example with discrete accelerometers. In [24], the authors proposed a quartz accelerometer featuring a multi-point temperature sensing (i.e., multi-thermometer), whose output signals were processed with neural networks algorithm. However in that work, the experimental data were collected using a temperature-controlled box, so it appears that thermal gradients could not be intentionally generated.

Following the paradigm of GAIN, we present a novel ML based approach for temperature rejection in an accelerometer featuring eleven thermometers. The use of ML relieved us from determining the mathematical relationship between acceleration and temperature. To acquire the required labelled data for training and testing, we specifically developed a training platform to generate them. This platform featured two setups to induce temperature and thermal gradients variations: the first used heating mats attached to the faces of the accelerometer box; the second used a lamp whose radiation heated up the accelerometer. These setups allowed us to demonstrate that, besides compensating for temperature, the ML algorithm was able to generalize over different measurement scenarios.

## 2. Experimental setup

The experimental setup consists of two parts: 1) a multi-sensor accelerometer, also referred to as GAIN, and 2) a training platform. The main difference is that GAIN is meant to be used during the operations, e.g., during a gravity survey, while, the training platform is used only in the lab to produce labelled data.

### 2.1. Multi-sensor accelerometer – GAIN

The multi-sensor system represents the first pillar of the GAIN paradigm and it features primary sensors and supplementary sensors. The primary sensors measure the quantity of interest (measurand) whereas the supplementary sensors measure the disturbances affecting the primary sensors. This concept could be applied to many different types of sensors, but in this work, the primary sensor was a tri-axial accelerometer and the supplementary sensors were eleven high precision thermometers.

The accelerometric sensors (one is shown in Fig. 1) were initially built as early prototypes of the Italian Spring Accelerometer (ISA), a payload of the European Space Agency (ESA) mission BepiColombo. As described in [16], the measuring principle was based on the mechanical oscillator and bridge capacitive transducer concept. The mechanical oscillator was milled from a single chunk of aluminum (denoted by the blue arrow in the figure) and weighed 450 g, along with the other components of the sensor. Their background noise spectral density was about $10^{-7} m/s^2/\sqrt{Hz}$ within the operational bandwidth of $\sim 5 \times 10^{-5}$ Hz to $\sim 0.05$ Hz. This was sufficient for our purposes and further improvements of the precision would not add value to our results. To measure all acceleration components, three sensors were arranged orthogonally inside an aluminum box (see Fig. 2 and Fig. 3), forming the tri-axial accelerometer. The sampling frequency was set to 20 Hz, where each record comprised the measures of all three components of the acceleration.

The eleven thermometers were made using PT10000 Platinum resistors in Wheatstone bridge circuits with high stability resistors (Vishay PTF56, 10 kΩ, 0.125 W). The power supply of the bridges was provided by one REF195 voltage reference, whose precision and thermal stability were ±5 mV and 5 ppm/°C respectively. The output voltages of ten Wheatstone bridges were acquired with a high-precision analog-to-digital converter (the ADS1263, featuring ten channels at 32bits) while the output of the remaining thermometer was acquired with a dedicated channel of the accelerometer acquisition electronics (at 24bit). The sampling frequency of each of the ten thermometers acquired by the ADS1263 was 1 Hz, while that of the thermometer acquired with the electronics of the accelerometer was 20 Hz. The thermometers were calibrated and characterized by placing them in close thermal contact with each other inside an oven in which the temperature varied by ~ 10 °C. Their precisions were better than $10^{-3}$ °C and the pair-wise Pearson correlations of their measures were better than 0.99986. After calibration and test were completed, the thermometers were displaced inside the accelerometer as shown in Fig. 2 (brown elements) and as indicated in Table 1.

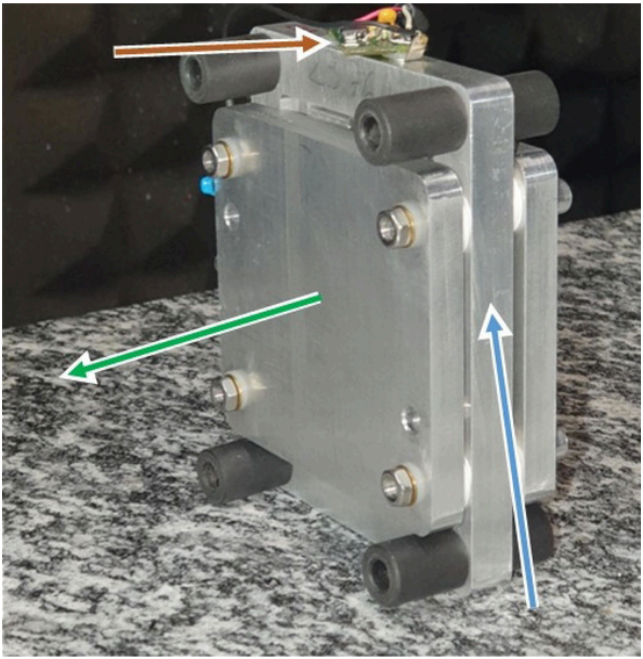

**Fig. 1.** An accelerometric sensor and its frontend electronics (indicated by the brown arrow) in which the sensitive axis is indicated by the green arrow. The three aluminum elements form two capacitors of the bridge transducer. The element indicated by the blue arrow consists of a proof mass attached with an elastic element to the external frame, thus forming the mechanical oscillator. (For interpretation of the references to color in this figure legend, the reader is referred to the web version of this article.)

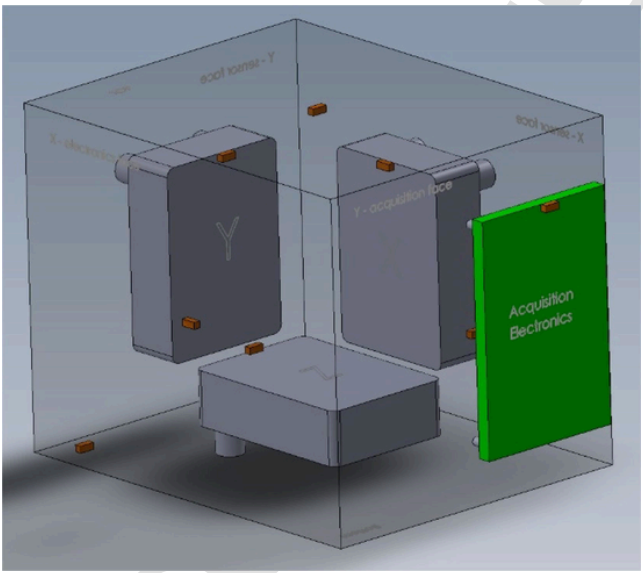

**Fig. 2.** Conceptual drawing of GAIN. The enclosing box is represented in transparency (it is actually made of aluminum, see Fig. 3 and Fig. 4) and some components, such as the cabling, are not represented. The three grey components represent the three accelerometric sensors (like that shown in Fig. 1), the tiny brown elements represent the thermometers, and the green board represents the electronics of GAIN (i.e., the ESP32, the ADS1263, the DPS310, etc.). (For interpretation of the references to color in this figure legend, the reader is referred to the web version of this article.)

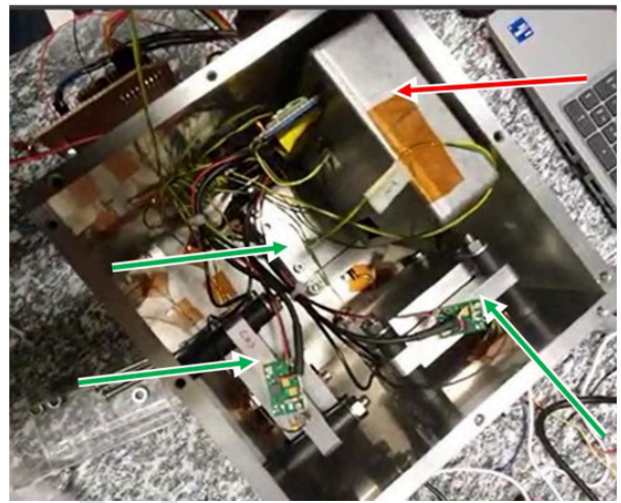

**Fig. 3.** This photo shows the interior of GAIN. The red arrow indicates the metallic box containing the accelerometer acquisition electronics (not shown in Fig. 2). The three accelerometric sensors (like that shown in Fig. 1) are indicated by the green arrows. The eleven PT10000 thermometers are connected via green and yellow wire pairs and secured with brown Kapton tape. (For interpretation of the references to color in this figure legend, the reader is referred to the web version of this article.)

Another supplementary sensor was a DPS310 MEMS barometer featuring an additional embedded thermometer. It was secured outside the accelerometer box and it had a precision of $\pm 5 \times 10^{-3}$ hPa and $\pm 0.5$ °C for pressure and temperature respectively. However, its measurements were not used during the data processing because they did not lead to any improvement of the results.

In order to correctly relate the measurements from all sensors, it was key to refer each reading to its own acquisition time. This was achieved with a dedicated acquisition system based on an ESP32 microcontroller featuring several data interfaces and a Wi-Fi connection. This device acquired the data from the sensors in real-time labelling each record with a time-stamp. In this way, it was easy to synchronize the records in post-processing. Finally, all acquired data were sent via Wi-Fi connection to a server for data storage.

**Table 1**
Locations of the twelve thermometers of GAIN with respect to the box containing the accelerometric sensors.

| Thermometer Name | Location |
| --- | --- |
| T z [C] | Sensor Z, next to the elastic element |
| T x [C] | Sensor X, next to the elastic element |
| T y [C] | Sensor Y, next to the elastic element |
| bottom x1 side [C] | Bottom face |
| bottom x2 side [C] | Bottom face |
| x sensor face [C] | Face where the X sensor is secured to, between to the securing screws |
| y sensor face [C] | Face where the Y sensor is secured to, between to the securing screws |
| x elect face [C] | Center of the face opposite to that of sensor X |
| y ACQ face [C] | Center of the face opposite to that of sensor Y |
| top [C] | Center of top face |
| inside air [C] | Suspended in the center of the box (not shown in Fig. 2) |
| external [C] | Embedded in the barometer outside the box (not used in data processing) |

## 2.2. Training platform

The training platform is the third pillar of the GAIN paradigm. In this work, we only considered disturbances related to the temperature. Therefore, the role of the training platform was to expose the accelerometer to all temperature variations and thermal gradients that might occur in the operative environment. To achieve this, we attached six heating mats to the six faces of GAIN (i.e., of the accelerometer box) as visible in Fig. 4. The heating mats dissipate heat when a voltage is applied to their input; changing this voltage, we could regulate the amount of heat. Each mat was activated with a relay controlled by a Raspberry Pi. The input voltage instead, was provided by a bench power supply whose output was also controlled by the Raspberry Pi.

The Raspberry Pi was programmed to create random temperature and thermal gradients fluctuations. This was accomplished by randomly activating a mat, or pair of mats, for a random duration (between a few minutes and 30 min), followed by a deactivation for another random duration (between few minutes and an hour). Additionally, the voltage supplied to the mats was also random between 1 and 12 V.

An additional setup to generate temperature variations was specifically implemented to collect testing data. In this case, the mats remained off while a lamp (red arrow in Fig. 4), with an output of approximately 260 W, was switched on for a random duration of time (between few minutes and 30 min). In addition, by time to time we manually moved the lamp to create different thermal gradients.

We emphasize that the described training platform is designed for applications where the accelerometer maintains a static configuration, such as in seismic monitoring, or consistently operates with a pose similar to that used during its training. For instance, the latter case is observed in airborne gravimetry when the aircraft can maintain a predetermined flight attitude. In cases where a dynamic signal is expected—e.g., where the accelerometer oscillates during measurements—the training platform must be capable of replicating such a signal as well.

## 3. Methods

### 3.1. Machine Learning

ML is the second pillar of the GAIN paradigm and plays a key role, as the proposed experiment relies on its ability to model complex systems directly from data (called training data). In the literature, many types of ML are described, such as supervised ML, unsupervised ML, and reinforcement learning. Supervised ML is the best suited for this work and means that the training data are labeled, i.e., they include the desired output (also called labels or target values). For example, in order to train a supervised ML algorithm to classify cat and dog pictures, the training data must be labeled pictures of cats and dogs. After training, the algorithm can generalize over new, unlabeled data and correctly predict the output (i.e., determine whether it is a cat or a dog). There are four steps in a typical ML project: data acquisition, data preparation, model building, and model deployment.

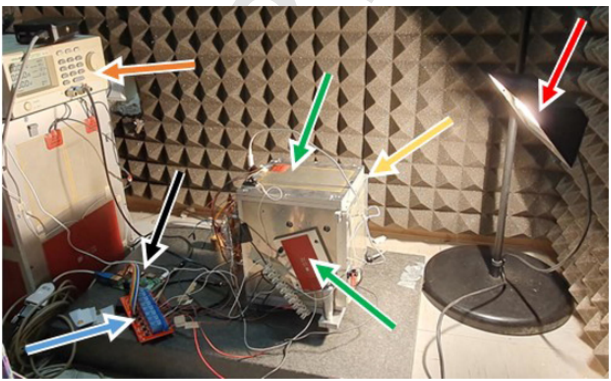

**Fig. 4.** Photo of the experimental setup where GAIN is indicated by the yellow arrow. Additionally, two (out of six) heating mats (green arrows), the lamp (red arrow), the relays (blue arrow), the power supply (orange arrow), and the Raspberry Pi (just barely visible; black arrow) are displayed. (For interpretation of the references to color in this figure legend, the reader is referred to the web version of this article.)

In this work, data were generated using the experimental setup described in the previous section whereas data acquisition methods are thoroughly described in Section 3.2.

Data preparation is described in Section 3.3 and consists of data pre-processing and feature engineering. Pre-processing means data cleaning and filtering, while feature engineering involves shaping the input in such a way that the ML can more easily model the system from the data.

Model building is an iterative process in which the ML algorithms are defined, trained, and tested. The term "model" is referred to a ML jargon according to which, training a ML algorithm yields a ML model. The ML algorithms used in this work are described in Section 3.4. During the training phase, a ML algorithm learns from the training data. To assess the trained model accuracy, some labelled data, called test data, are excluded from the training process. Thus, applying the trained model to them, we have both the predicted values and the desired output. Comparing them provides an estimate of the ML model accuracy. Summary statistics such as Root Mean Squared Error (RMSE), $R^2$ (a.k.a., coefficient of determination), Mean Average Percentage Error (MAPE) as well as graphical tools such as time plots, hexagonal binning plots, and histograms, are typically used to assess the performance of the ML models and select the best option. The reader is referred to the webpage of the Python package Scikit-Learn (scikit-learn.org) and to [4] for a detailed description of these summary statistics and graphical tools. In this context, it is just important to know that, ideally (for a model able to provide exact predictions), RMSE and MAPE should be zero while $R^2$ should be one. Additionally, we calculated a custom summary statistic identified as STD RR (which is a short for standard deviation rejection ratio). This was defined as the following ratio:

$$STD\ RR = \frac{std\ (desired\ output)}{std\ (error)}$$

where *std* is a function yielding the standard deviation, *desired output* is the timeseries of target values, *error* is the timeseries of the deviations of the predicted values from the target ones. Notice that STD RR is a rejection ratio, which means that the higher it is, the better the temperature effects are rejected from the accelerometer output.

The outcome of the model building enables the deployment of the ML model in the operational phase. For instance, our application uses the ML model to eliminate disturbances from airborne gravity measurements. Deployment is not covered in this paper as we focused only on the proof of the presented principle to remove disturbances from the measurements.

### 3.2. Data acquisition

For the sake of clarity, we recall that the output of an accelerometer perfectly reflects the measured acceleration only in the ideal case; the output of a real accelerometer instead combines the effects of the acceleration to those of the disturbances, such as the temperature. In the hardware setup outlined in Section 2.2, the accelerometer was stationary and the local micro-seismic signals were negligible: thus, the acceleration was constant. Nevertheless, the accelerometer output did not remain constant due to the temperature fluctuations caused by the heating mats (or the lamp). Thus, the accelerometer output was almost uniquely reflecting the temperature variations, which are the disturbances that need to be compensated for.

Given such conditions, we used the accelerometer output as the desired output of the ML model, and the temperature readings from ther-mometers as input. Hence, the ML model was trained to predict the effect of the temperature, i.e., the error, on the accelerometer output (see Fig. 5).

After the ML model has been trained, it can be deployed in the operational context. During this phase, the accelerations are not negligible therefore the accelerometer output combines both the accelerations and the errors due to the temperature variations. However, the ML model is able to predict the error which can be subtracted from the accelerometer output yielding the measurements of the acceleration (see Fig. 6).

Notice that this method for deployment only works in linear regime. In other words, when the input accelerations are comparable to the gravity, non-linear effects might significantly affect the performance of the ML model. Such cases were not covered in this work.

Finally, routine data acquisition and training of the ML model are not required unless the hardware is modified.

### 3.3. Data preparation

Each sensor of GAIN had its own sampling rate, therefore all records had to be registered to each other in order to be referred to the same timing. To do so, we interpolated and oversampled the temperatures to the sampling frequency of the accelerometer (20 Hz). Then, we synchronized the records obtaining a timeseries where each entry comprised readings from all sensors (e.g., this can be done using the function *merge_asof* of Python-Pandas library).

We filtered the data with a band-pass, second order, Butterworth filter with cutoff frequencies of $4.63 \times 10^{-5}$ Hz (6 h period) and 0.05 Hz (20 s period). The goal was to remove signals unrelated to the tempera-ture, such as instrumental drift and noise, while preserving those of interest, such as the gravity variations. Consequently, we chose the lowest frequency in order to dump the very slow drift of the instrument without affecting the gravity signals. In fact, the period of the latter is typically shorter than six hours in airborne surveys because of the duration of the flight. On the other side, filtering the high frequencies seemed unnecessary because the thermal capacity of the accelerometer already acted as a low-pass filter for temperature. Anyway, we wanted to remove signals such as sensor noise, local micro-seismic signals, anthropic noise, etc., that were unrelated to temperature and may have affected the ML performance. Finally, although the cutoff frequencies might have required further fine-tuning, the values we used were satisfactory for our proof-of-concept experiment. After filtering, the 20 Hz sampling frequency was unnecessary; therefore, we down-sampled the time series to 1 Hz.

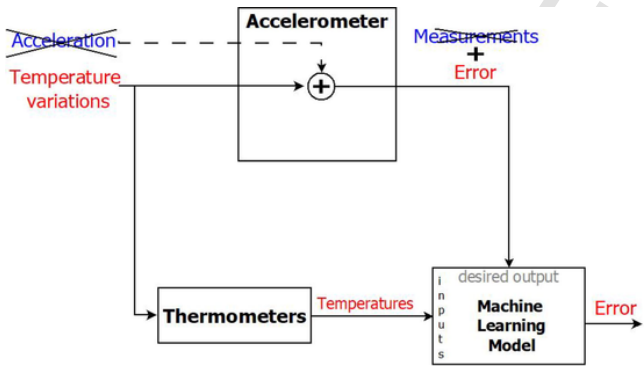

**Fig. 5.** Schematic representation of the labelled data acquisition procedure. As the accelerometer was stationary, its output was only due to the temperature variations. In other words, the accelerometers output was the error caused by the temperature variations. Therefore, the ML algorithm was trained to reconstruct from the thermometers output the error caused by the temperature.

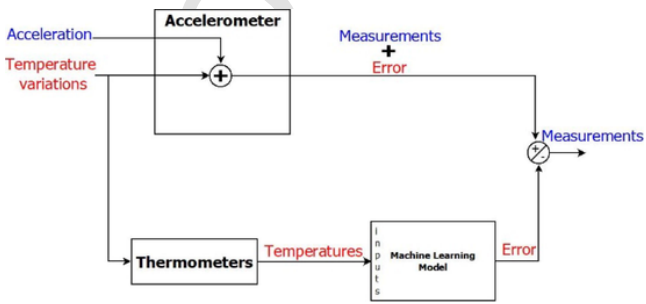

**Fig. 6.** Schematic representation of the data acquisition and processing during the operational phase. In this case, the accelerometer output combines acceleration measurements and errors due to temperature variations. The ML model can predict this temperature-induced error which can be subtracted from the accelerometer output yielding the measurement.

As the data acquisition lasted for more than two months, the accelerometer recorded events unrelated to the temperature but not negligible as assumed in Section 3.2. Examples included anthropic activities near the instrumentation, earthquakes, and more. These events were not completely dumped by the band-pass filter and could potentially have affected the training process; thus, we dropped them out from the data.

In ML, it is essential to test the ML model against fresh data that were not used during the training [8]. Complex algorithms, such as neural networks, require also validation data to assess the progress of the training process. For this reason, the timeseries were split into three subsets: training, validation, and testing. Validation set were merged with the training set when training non-neural networks algorithms (i.e., when validation data were not required).

All sensors were calibrated to yield measurements in the conventional measurement units (m/$s^2$, °C, etc.), allowing for easy functional checks. For instance, thermometer outputs above or below the expected range would be considered as a warning of some bug or malfunctioning. However, in general ML algorithms do not work well if their input signals (also called "features") vary over a wide range [8]. For instance, if the temperature ranges from 10 to 20 °C and the pressure ranges from 900 to 1100 mbar, the ML algorithm may have difficulty converging to a good ML model. For this reason, it is common practice to normalize each feature prior to feeding them into the algorithm, so that their standard deviation is one and their average is zero. The coefficients to normalize the features must be computed over the training data and then applied to validation and testing sets.

For preparing the input data for the ML algorithms, we utilized three approaches: instantaneous values, time derivatives, and time intervals. Instantaneous values were the measurements taken from all sensors at the same time. Time derivatives were derived by computing the ratio between the differences between two consecutive measurements and the sampling period. Time intervals were short timeseries of temperature values which may be arranged into two-dimensional matrices and visualized as a one-channel images (like grey level images), as shown in Fig. 7. Each row of an image was associated to a thermometer, each column was associated to a recording time, and the level of each pixel to a temperature value (after filtering). The time interval covered by one matrix was 240 s and consecutive columns were spaced 10 s apart, resulting in images with 11x25pixels.

The rationale behind using time derivatives and time intervals as inputs of the ML algorithm was to provide it also with information related the time evolution of the temperatures rather than just their instantaneous values. Such information should enable the algorithm to better modelling the system and providing more accurate predictions.

### 3.4. Model building

Predicting a continuous variable, such as an acceleration, from one or multiple inputs is called regression. There are a number of different

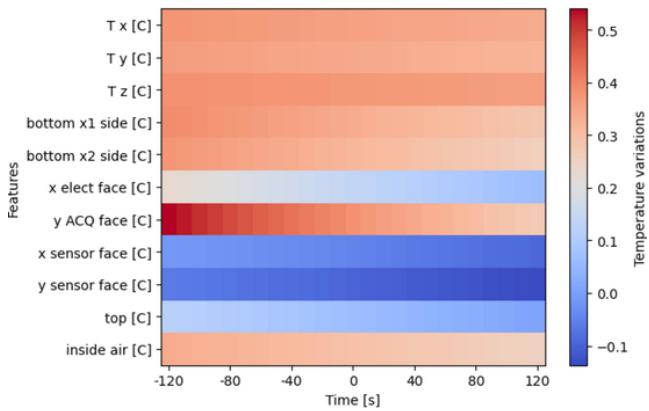

**Fig. 7.** Image representing the time interval of the temperatures fed to the ML algorithm. The color scale representing the temperature values is shown on the left.

ML algorithms for regression, but the most widely used is Linear Regression – i.e., a linear combination of the inputs. Although it is relatively simple, it is effective in many applications where non-linear effects are negligible. On the other hand, the most powerful algorithms currently known are the Neural Networks, which can accomplish incredibly complex tasks [2]. A typical Neural Network architecture is the Feed-Forward Neural-Network (FFNN), which consists of layers of neurons where the information propagates from an input layer, through hidden layers, to an output layer. When the network has many hidden layers, it is said "deep", hence the so called Deep Neural Networks which are the foundation of the Deep Learning. Usually, deeper networks are able to solve more complex tasks but at the price of greater complexity and difficulty in tuning multiple hyperparameters such as: number of layers, number of neurons, activation functions, etc. [8].

In this proof of principle project, we tested and compared the Linear Regression and FFNN algorithms described in Table 2. We distinguished two types of Linear Regression: the simple version, accepting only one input signal; and the multivariate version, accepting multiple input signals. The first one represents the conventional method for temperature rejection, where only one thermometer is used. Instead, the second complies with the GAIN paradigm, where multiple thermometers are used. Finally, each component of the accelerometer was addressed separately obtaining one ML model for each of them.

Many programming languages can be used to implement ML algorithms; in this project we used Python and its specific modules for conventional algorithms (scikit-learn, https://www.scikit-learn.org) and neural networks (Keras, https://www.keras.io).

## 4. Results

### 4.1. Exploratory analysis of labelled data

Experimental data [14] were acquired using the experimental setup and the acquisition method described in Sections 2 and 3.2 respectively, resulting in the four datasets outlined in Table 3. Datasets A1 and A2 consisted of instantaneous values along with their time derivatives, while B1 and B2 consisted of time intervals and were obtained from the first two respectively by resampling, as explained in Section 3.3. The testing subsets of A1 and B1 were acquired using the experimental setup involving the lamp (from here, lamp-setup), while those of A2 and B2 were acquired with the setup containing the heating mats (from here, mats-setup).

All datasets included the measures of the three acceleration components (x, y, and z), which served as the desired output (label). However, the ML algorithms were trained to predict only one component at a time. Specifically, we used either the vertical component z or the horizontal component x, and did not report results for the y component, as they were equivalent to those for the x.

The plots in Fig. 8 show the acceleration components of the whole dataset (i.e., dataset A1), which was collected within a period of about 2.5 months. The gaps about the 15th of February and the 1st of March were due to technical stops. The red vertical lines separate the data collected using the mats-setup from those collected with the lamp-setup. Datasets A2 and B2 comprised only data obtained with the mats-setup from 2nd of February to the 15th of March.

Fig. 9 shows a single day of the entire dataset, allowing for closer inspection. The top plot displays the accelerometer outputs, while the bottom plot shows temperature measurements. Each peak in the temperature plot corresponds to an activation of a heating mat (or a pair of heating mats), and usually produced a peak in the accelerometer output. As instances, two peaks have been annotated in the temperature plot, indicating which mats were activated.

### 4.2. Machine Learning performance

The ML algorithms specified in Section 3.4 were trained and tested using the datasets listed in Table 3 and the summary statistics (see Section 3.1) of the results are reported in Table 4. The best performing algorithms for each dataset/component are reported in bold. The train-ing phase took from few seconds for linear regression up to few minutes for FFNN B (100 epochs) on a Dell Poweredge 730, 2 CPU Xeon, 512 Gb RAM with GPU Nvidia Tesla K80.

**Table 2**
ML algorithms description. For the FFNNs, the hyperparameters name are referred to the Keras syntax and a clear definition can be found in [8].

| Algorithm | Input | Hyper-parameters |
|---|---|---|
| Linear Regression | Instantaneous 1 thermometer | None |
| Multivariate Linear Regression | Instantaneous + Time derivatives 11 thermometers | None |
| FFNN A | Instantaneous + Time derivatives 11 thermometers | **Layers**: Flattens, Dense(12), Dense (12), Dense(1); **Loss Function**: Mean Absolute Error **Activation function**: elu; **kernel_regularizer**: l2(0.1) |
| FFNN B | Time interval 11 thermometers | **Layers**: Flattens, Dense(12), Dense(1); **Activation function**: LeakyReLU (alpha = 0.2); **Loss Function**: Mean Absolute Error |

**Table 3**
Datasets used for training and testing the ML algorithms.

| Dataset name | # of features | # records | # training records | # validation records | # testing records | Exper. setup for training data | Exper. setup for testing data |
|---|---|---|---|---|---|---|---|
| A1 | 22 (11 instant. values + 11 time derivatives) | $4.5 \times 10^6$ 100 % | $2.51 \times 10^6$ 55.8 % | $0.28 \times 10^6$ 6.2 % | $1.71 \times 10^6$ 38 % | Mats-setup | Lamp-setup |
| B1 | 11 × 25pixels | $4.5 \times 10^5$ 100 % | $2.51 \times 10^5$ 55.8 % | $0.28 \times 10^5$ 6.2 % | $1.71 \times 10^5$ 38 % | Mats-setup | Lamp-setup |
| A2 | 22 (11 instant. values + 11 time derivatives) | $2.0 \times 10^6$ 100 % | $1.35 \times 10^6$ 67.5 % | $0.15 \times 10^6$ 7.5 % | $0.5 \times 10^6$ 25 % | Mats-setup | Mats-setup |
| B2 | 11 × 25pixels | $2.0 \times 10^5$ 100 % | $1.35 \times 10^5$ 67.5 % | $0.15 \times 10^5$ 7.5 % | $0.5 \times 10^5$ 25 % | Mats-setup | Mats-setup |

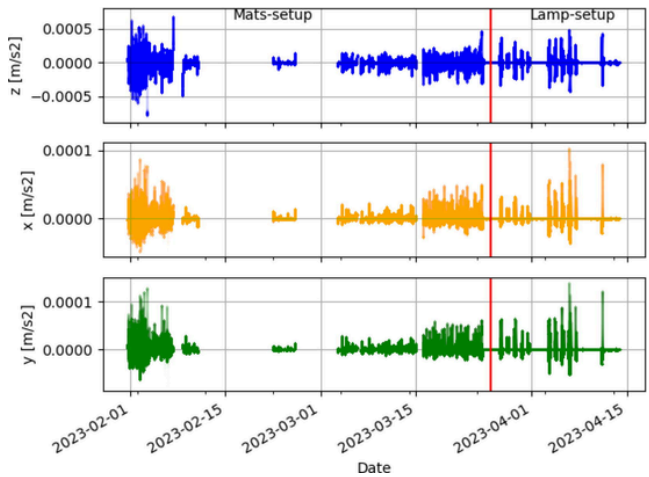

**Fig. 8.** Time plots of the acceleration components measured in m/s².

Fig. 10 and Fig. 11 show the time plots (an interval of few hours of April 5th) of the results obtained using the algorithm FFNN A with the dataset A to predict the components Z and X respectively. The hexagonal binning plots of Fig. 12 show the results obtained for the X, rows 6 to 9 of Table 4. The histograms in Fig. 13 show the error distribution obtained using the configurations outlined in rows 6 to 9 of Table 4.

The plot in Fig. 14 shows the performance (specifically the coefficient of determination $R^2$) of multivariate linear models with increasing number of input features (i.e., of thermometers). Here, the predicted variable was the x component of the acceleration. For instance, to make the blue line, we first trained a linear regression model using only "Tx [C]", then using "Tx [C]" plus "bottom X1 side [C]", then "Tx [C]" plus "bottom X1 side [C]" plus "bottom X2 side [C]", and so on in the order indicated by the "Cumulative features" axis at the bottom. In other words, the x-axis represents the cumulative features added to the model. Thus, each data point corresponds to the inclusion of a specific feature along with all the previous features. Similarly, we did for the red line where the Cumulative features are in reversed order as indicated at the top. The meaning of displaying the Cumulative features in reversed order was to show that the contribution of each feature might depend on those included in the model before. For example, adding the thermometer "y sensor face [C]" in the blue curve was beneficial with $R^2$ increasing of almost 0.2, instead in the red curve $R^2$ worsened of the same amount.

## 5. Discussion

This study provides insights into how temperature variations affect the output of a high-accuracy accelerometer. As anticipated in the introduction, the temperature impacts the sensitivity by modulating the output. This explains the greater variations of the vertical component (z) of the accelerometer compared to the others (see top plot of Fig. 9). In fact, the output of the z component resulted from the modulation of gravity (to which it is aligned) by temperature. On the other hand, the accelerations measured by the horizontal components were zero because they were oriented orthogonally to the gravity. Thus, their modulation by the temperature was insignificant. This was further demonstrated by the very small coefficient of determination ($R^2$ = -0.006) obtained when fitting the x component with the temperature of the accelerometric sensor x, measured by the thermometer T x [C] (see Table 4, row 5).

A closer examination of the data revealed that the temperature affected the outputs through an additional phenomenon. Specifically, we observed that the x and y components were mostly affected by the temperatures of the box faces. This is visible from Fig. 9, where components x and y exhibited a correlation with the temperatures of the box faces (i.e., x sensor face [C], y sensor face [C], etc.). Fig. 14 provides further evidences for the component x leading to the same conclusion. As suggested by [19], the warpage of the accelerometer box can likely explain these results. In other words, the temperature induces a variation of the accelerometer output by slightly changing the shape of its box and, thus, the alignment of the sensors with respect to the gravity. It is important to notice that all thermometers were key to reconstruct the component x, while using only one thermometer (e.g., X sensor face [C]) would lead to unsatisfactory results (see Table 4, row 6). This was clearly proven also by the results shown in Fig. 14 as the best performance was achieved when all thermometers were used. We therefore concluded that measuring the temperature in multiple spots, as recommended by [24], was critical for accurately removing its effect. In other words, beside temperature, also the thermal gradients affected the out-put of an accelerometer and compensating for their effects required multiple thermometers properly displaced.

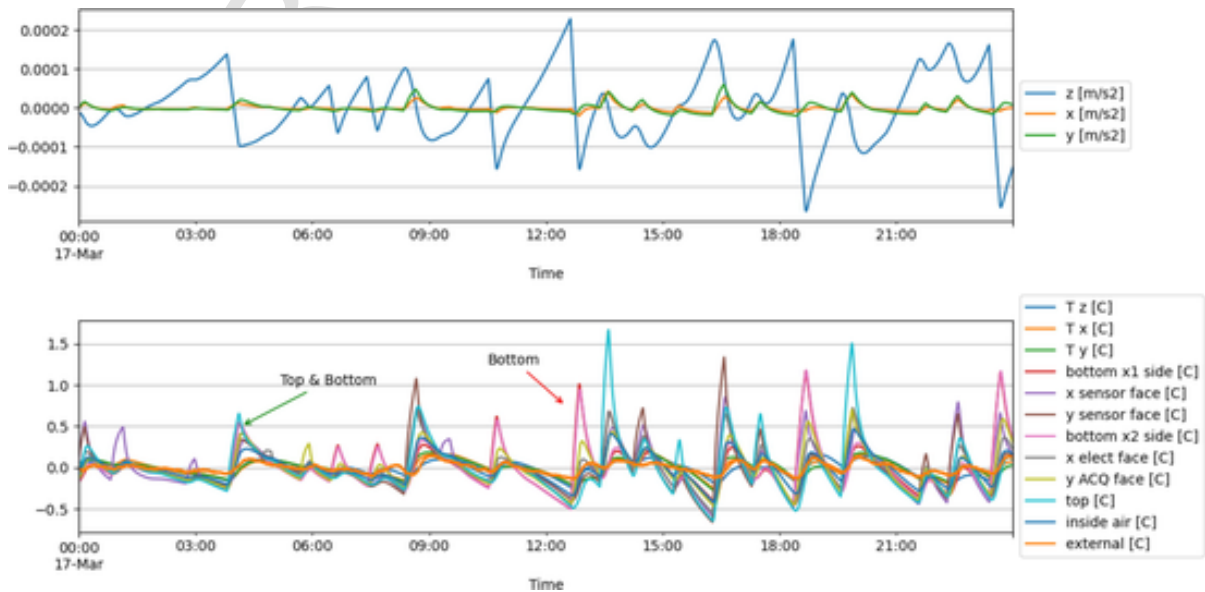

**Fig. 9.** The time plot at the top shows the acceleration components (measured in m/s²), the time plot at the bottom the temperatures (measured in °C) after filtering.

**Table 4**
Summary statistics of the ML performances. The thermometer used for the Linear Regression is indicated next to the dataset name.

| # | Accel. Comp. | Algorithm | Data set | Results | | | |
|---|---|---|---|---|---|---|---|
| | | | | STD RR | RMSE [m/s$^2$] | R$^2$ | MAPE |
| 1 | Z | Linear Regression | A1 - T z [C] | 4.4 | $1.5 \times 10^{-5}$ | 0.947 | 2.8 |
| 2 | | Multivariate Linear Regression | A1 | 25.5 | $2.7 \times 10^{-6}$ | 0.998 | 4.2 |
| 3 | | **FFNN A** | **A1** | **27.2** | $2.5 \times 10^{-6}$ | **0.999** | **2.2** |
| 4 | | FFNN B | B1 | 18.7 | $3.8 \times 10^{-6}$ | 0.997 | 1.3 |
| 5 | X | Linear Regression | A1 – T x [C] | 1.0 | $6.6 \times 10^{-6}$ | -0.006 | 3.0 |
| 6 | | Linear Regression | A1 - X sensor face [C] | 1.0 | $5.1 \times 10^{-6}$ | 0.391 | 6.2 |
| 7 | | Multivariate Linear Regression | A1 | 3.2 | $2.1 \times 10^{-6}$ | 0.901 | 7.6 |
| 8 | | FFNN A | A1 | 3.0 | $2.2 \times 10^{-6}$ | 0.888 | 3.9 |
| 9 | | **FFNN B** | **B1** | **3.4** | $2.0 \times 10^{-6}$ | **0.912** | **2.6** |
| 10 | Z | Linear Regression | A2 - T z [C] | 2.0 | $2.1 \times 10^{-5}$ | 0.757 | 2.6 |
| 11 | | Multivariate Linear Regression | A2 | 6.8 | $6.3 \times 10^{-6}$ | 0.978 | 0.8 |
| 12 | | FFNN A | A2 | 8.5 | $5.1 \times 10^{-6}$ | 0.986 | 0.6 |
| 13 | | **FFNN B** | **B2** | **14.4** | $3.0 \times 10^{-6}$ | **0.995** | **0.4** |

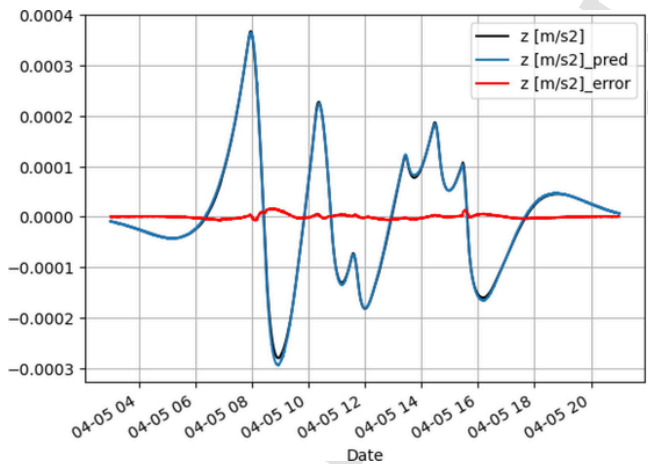

**Fig. 10.** Time plots of desired output (z [m/s2]), predicted output (z [m/s2]_pred) and their difference (i.e., the error, z [m/s2]_error). This result was taken from the dataset A1 and obtained using the algorithm FFNN A.

When inspecting the time plots in Fig. 9, it appeared challenging to determine a mathematical relationship between the accelerometer output and the temperatures. This underscored the need of employing Machine Learning techniques to uncover such relationships. However, successfully training a ML algorithm required labeled data that comprehensively represented the operational environment. Hence, the development of a training platform was imperative. It is noteworthy that, in contrast to setups reliant on thermal chambers like the one employed by [24], our training platform offered the unique capability of intentionally inducing a thermal gradients.

We conducted a comparative analysis of two distinct ML algorithms: linear regression and FFNN. The findings indicated that the relationships between temperature and accelerometer outputs were predominantly linear. Consequently, employing FFNN did not yield significant enhancements, and the efforts invested in tuning hyperparameters appeared unjustified. Linear regression, on the other hand, offers the advantage of easily interpretable results. Indeed, the resulting parameters, specifically the coefficients of the regression, allow straightforward interpretation. For instance, a temperature with a small coefficient is likely to have less impact on the accelerometer compared to one with a large coefficient. This interpretability can be leveraged in experimental development to pinpoint the most effective thermometer placements. However, it is noteworthy that FFNN consistently outperformed linear regression in all instances, indicating its superiority. This superiority becomes especially critical when exploring the effects of other types of disturbances, such as unaccounted rotations of the reference frame, which involve non-linear interactions.

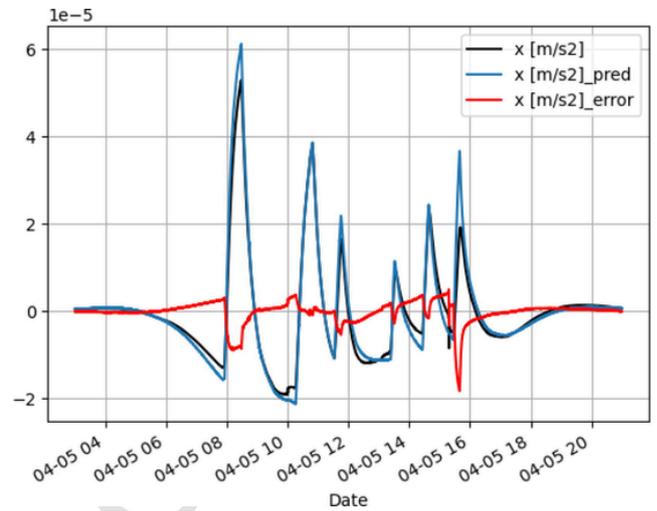

**Fig. 11.** Time plots of desired output (x [m/s2]), predicted output (x [m/s2]_pred) and their difference (i.e., the error, x [m/s2]_error). This result was taken from the dataset A1 and obtained using the algorithm FFNN A.

One of the most important capabilities of ML models is their ability to generalize over (i.e., being effective on) new data. This capability must be proven before the model is deployed. Usually, this is achieved by testing the model against labeled data that were never used during training. In this work, we went beyond this concept by testing the model on testing data acquired using a different experimental setup (i.e., the lamp-setup). The purpose of this experiment was to simulate a scenario in which the training data are collected on the training platform but the system is then deployed on the field, where the temperature variations are driven by a completely different source. The results in rows 10 to 13 of Table 4, obtained using the same setup (mats-setup) for both training and testing data, were reported for comparison. In this case, the FFNNs (row 12 and 13) outperformed the multi-variate linear regression (row 11). This did not happen to the same extent (e.g., see rows 8 and 9) when testing data and training data were acquired using the two different setups (mats-setup for training and lamp-setup for testing). One explanation could be that the FFNNs were able to learn specific characteristics of the whole experimental setup, including the behavior of the heating mats. However, these characteristics were not useful when the lamp-setup was used, therefore the FFNNs performed similarly to the multi-variate linear regression. In other words, the FFNNs were overfitting the experimental setup.

## 6. Conclusion

This study presents a novel approach to compensate for temperature effects on high sensitivity accelerometers. The GAIN paradigm was utilized, employing multi-sensor techniques with Machine Learning, and a dedicated platform for collecting labelled data. Specifically, this study explored a multi-thermometer solution that has been underexplored in previous studies. We developed a prototype of a novel accelerometer with eleven thermometers and used machine learning to establish the relationship between temperature measures and accelerometer output. We created a training platform that replicates the operational thermal environment in a laboratory setting to obtain labeled data for ML training. This approach circumvents the need for costly and time-consuming field operations. Our platform differed from conventional thermal chambers typically used for calibrating accelerometer thermal behavior; in fact, it possessed the capability to intentionally generate random thermal gradients. The multi-thermometer approach enabled the detection of such gradients allowing us to demonstrate that their effect was significant and must be considered when high accuracy is required.

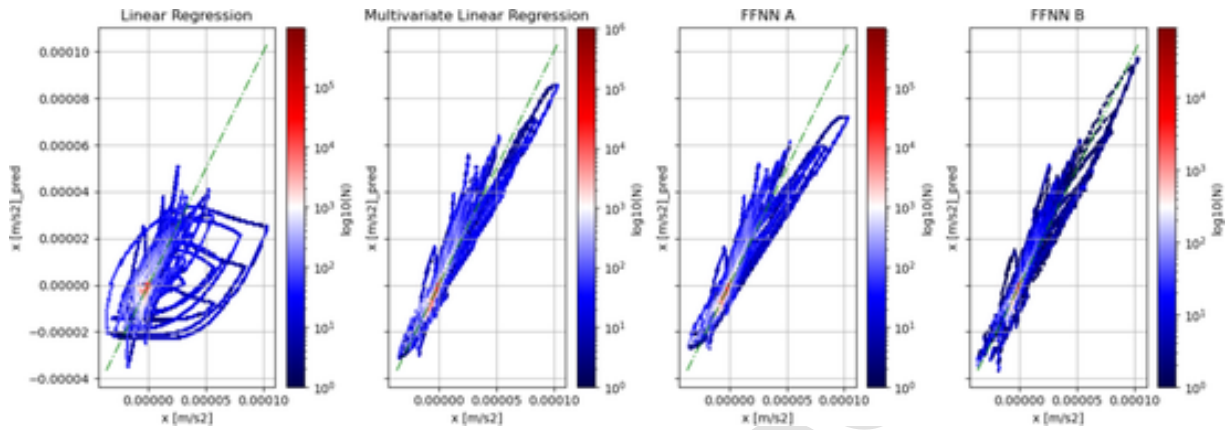

**Fig. 12.** Hexagonal binning plots of predicted output vs desired output. This kind of plots are similar to scatter plots in that they both display the relationship between two variables, but hexagonal binning plots are better suited for visualizing large datasets and displaying the number N of the data points. These results were obtained for the component X using the configurations outlined in rows 6–9 of Table 4. The green, dashed lines show where the points should ideally distribute. In other words, accurate predictions lie close to the dashed lines. (For interpretation of the references to color in this figure legend, the reader is referred to the web version of this article.)

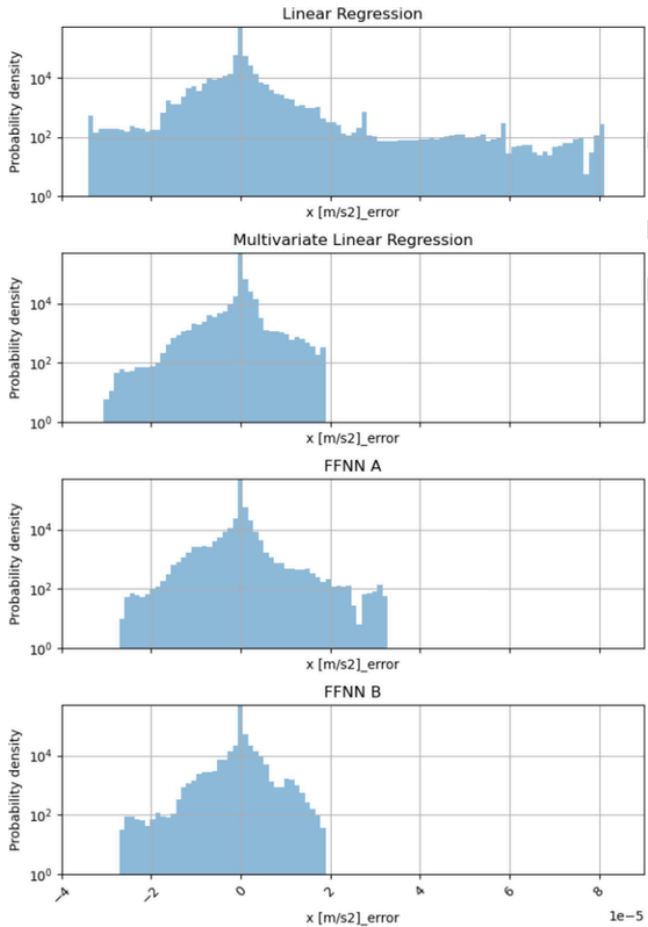

**Fig. 13.** Histograms of the errors associated to the predicted values of the component–using the datasets A1 and B1 (i.e., rows 6 to 9 of Table 4). The probability density is shown in logarithmic scale, the acceleration (x-axis) is measured in m/s$^2$. Notice that the x-axes of the four histograms have the same scale, indicated in the latter.

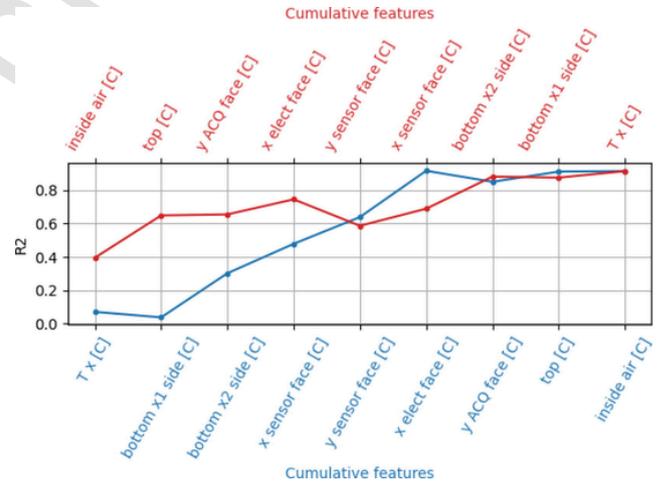

**Fig. 14.** The plot shows $R^2$ as function of the number of features (i.e., of thermometers) used to compose the input of the ML algorithm. Here, the predicted variable was the x component of the acceleration. For the blue curve, the features were added as listed at the bottom in blue (with the label "Cumulative features"). Similarly, the cumulative features for the red curve are listed at the top in red. (For interpretation of the references to color in this figure legend, the reader is referred to the web version of this article.)

Combining the multi-sensor approach with Machine Learning was key to achieving better results. To thoroughly test the capability to generalize of the ML models, we developed an experimental setup, featuring a heating lamp, specifically to generate testing data. Finally, the experimental results showed that the multi-thermometer approach, compared to the conventional method using a single thermometer, can improve the temperature rejection ratio (STD RR) by a factor up to 6.2 in the best case (this is the ratio between the STD RRs of rows 3 and 1 in Table 4).

We believe that our achievements are just a promising first step towards a more general application of the GAIN paradigm. Further im-provements can be made by determining the most effective positioning of thermometers, such as next to the feet of the box, the screws used to secure the sensors, or next to the acquisition electronics. Further experimentation should be done with non-static experimental setups for investigating the non-linear effects. This requires a more complex hardware, and the true acceleration must always be known for generating labeled data. The GAIN paradigm can also be applied to other types of disturbances, such as unaccounted rotations of the reference frame. However, this requires other supplementary sensors, like a gyroscope, and the training platform must apply rotations to the accelerometer. In such cases, the use of non-linear algorithms like an FFNN will be crucial. Overall, this study provides for valuable insights and techniques for improving accelerometer accuracy, which can be applied in various fields such as airborne gravimetry and aerospace. Finally, the same paradigm could be applied to other types of accelerometric sensors, such as MEMS devices, as well as different types of measurement systems, such as magnetometers and gyroscopes.

# 7. Declarations

## 7.1. Experimental data

Experimental data are publicly available here: https://data.mendeley.com/datasets/f78bmhr628/1.

# 8. Declaration of Generative AI and AI-assisted technologies in the writing process

During the preparation of this work the authors used ChatGPT 3.5 of OpenAI in order to improve language and readability of the text. After using this tool/service, the authors reviewed and edited the content as needed and take full responsibility for the content of the publication.

## CRediT authorship contribution statement

**Lorenzo Iafolla:** Conceptualization, Data curation, Formal analysis, Funding acquisition, Investigation, Methodology, Resources, Software, Validation, Visualization, Writing – original draft, Writing – review & editing. **Francesco Santoli:** Conceptualization, Funding acquisition, Methodology, Project administration, Resources, Supervision, Validation, Writing – review & editing. **Roberto Carluccio:** Methodology, Resources, Supervision, Validation, Writing – review & editing. **Stefano Chiappini:** Data curation, Methodology, Resources, Validation, Writing – review & editing. **Emiliano Fiorenza:** Investigation, Methodology, Supervision, Validation. **Carlo Lefevre:** Methodology, Supervision, Validation, Writing – review & editing. **Pasqualino Loffredo:** Investigation, Resources, Validation. **Marco Lucente:** . **Alfredo Morbidini:** Investigation, Resources, Validation. **Alessandro Pignatelli:** Formal analysis, Methodology, Supervision, Validation, Writing – review & editing. **Massimo Chiappini:** Conceptualization, Funding acquisition, Methodology, Project administration, Resources, Supervision, Validation, Writing – review & editing.

## Declaration of competing interest

The authors declare the following financial interests/personal relationships which may be considered as potential competing interests: Lorenzo Iafolla, Francesco Santoli, Roberto Carluccio, Stefano Chiappini, Emiliano Fiorenza, Marco Lucente, Massimo Chiappini have patent pending to INGV and INAF.

## Acknowledgements

This work was funded by "Regione Lazio" (Italy) with European Regional Development Fund (Italy, Lazio) through the call "Gruppi di Ricerca 2020 (POR FESR LAZIO 2014 – 2020), project number: A0375-2020-36674